\documentclass[pdftex,a4paper]{jpconf}
\usepackage{graphicx}
\usepackage{ifpdf}
\usepackage{amssymb}
\usepackage{lineno}
\begin{document}
\title{A method to rescale experimental data with dependence on $Q^2$ for DVCS process}

\author{Giacinto Ciappetta}

\address{Department of Physics, University of Calabria}
\address{Istituto Nazionale di Fisica Nucleare, Gruppo collegato di Cosenza \\ I-87036 Rende, Cosenza, Italy}

%\ead{thorsten.i.renk@jyu.fi}

%\author[label1,label2]{Giacinto Ciappetta}

%\address[label1]{Department of Physics, University of Calabria}
%\address[label2]{Istituto Nazionale di Fisica Nucleare, Gruppo collegato di Cosenza \\ I-87036 Rende, Cosenza, Italy}

\ead{giacinto.ciappetta@cs.infn.it}

\begin{abstract}
    We investigate the procedure for rescaling the DVCS cross section data %between two data sets
    collected with different invariant mass, $W$, of the virtual photon - proton system. We suggest a method which makes the rescaling more functional to conduct statistical analysis on overall data. The study can be applied to rescale data collected with different photon virtuality $Q^2$. Also we show a dependence on $Q^2$ for the $\delta$ parameter, that is used to describe the cross section as a function of $W$.
    \\
    \\
    DVCS; Rescaling
\end{abstract}

%\begin{keyword}

%DVCS \sep Rescaling %\sep
%{PACS numbers}: 13.60.Fz

%\end{keyword}

%
%\begin{PACS}
%      {13.60.Fz}{Elastic and Compton scattering}   \and
%      {14.20.Dh}{Protons and neutrons}
%     } % end of PACS codes
%\end{PACS}
%

% \linenumbers
%\linenumbers

%\section{}
%\label{}

%
\section{\label{intro}Introduction}
\begin{figure}
    \centering
    \includegraphics[width=\columnwidth]{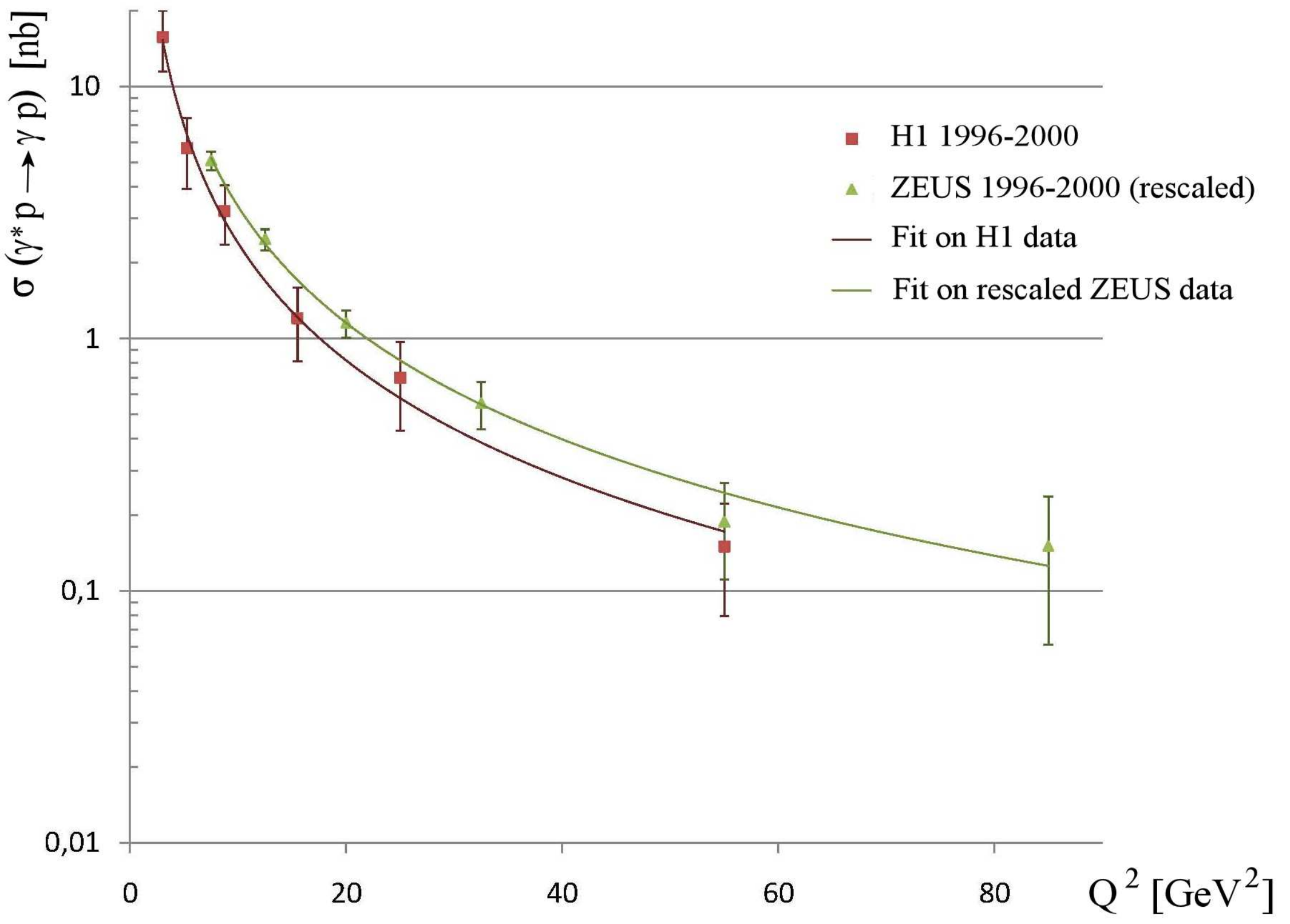}
    \caption{\label{fig:Ciappetta_DVCS2011-1_fig1}
%(Color on the Web only) DVCS cross section $\sigma(\gamma^\ast p \longrightarrow \gamma p$) as function of $Q^2$ for $W=82\,\textrm{GeV}$ ($|t| < 1.0 \,\textrm{GeV}^2$, where $t$ is the four momentum transfer squared at the proton vertex). The error bars re\-present the statistical and systematic uncertainties added in quadrature. The experimental data collected by the ZEUS Collaboration \cite{Chekanov:2003ya} have been rescaled to those collected by the H1 Collaboration \cite{Aktas:2005ty} applying Eq.~(\ref{eq:varepsilon-W}), where $\varepsilon_{_{W}} \simeq 0.9389$.
}
\end{figure}

The models used %proposed %\hspace{0.01mm}
to interpret the Deeply Virtual Compton Scattering (DVCS) process, $\gamma^\ast\textrm{p} \longrightarrow \gamma\textrm{p}$, are compared with an overall set of experimental results. This is obtained from several data sets collected at diffe\-rent energies.
The DVCS is the diffractive scattering of a virtual photon ($\gamma^\ast$) off a proton ($\textrm{p}$), i.e. $\gamma^\ast\textrm{p} \longrightarrow \gamma\textrm{p}$ where $\gamma$ denotes the outgoing photon.
%In particular, t
The integrated cross section can be written \cite{Aktas:2005ty} as a simple function: %of $W$, the invariant mass of the $\gamma^\ast\textrm{p}$ system, and $Q^2$, the virtuality of the photon:
\begin{equation}
\label{eq:dipendenza-Wdelta-Q2}
	\sigma%_{_{DVCS}}
                         (Q^2,W) \propto W^{\delta} \times \left( \frac{1}{Q^2} \right)^{n} \,,
\end{equation}
where $W$ is the invariant mass of the $\gamma^\ast\textrm{p}$ system and $Q^2$ is the virtuality of the photon.
%where
$\delta$ and $n$ are parameters obtained from fits to experimental data, % and
by keeping fixed respectively the $Q^2$-value or the $W$-value.
%%%%%%%%%%
The overall set of experimental results is given by a procedure which rescales the DVCS cross section measurements from an experiment to the kinematics of another experiment. %,
This is possible %
by applying some factors that are used also for the error analysis.
%%%%%%%%%%
As example, the ZEUS Collaboration \cite{Chekanov:2003ya} data, taken at $W = 89\,\textrm{GeV}$ and $Q^2 = 9.6\,\textrm{GeV}^2$, are rescaled to the H1 Collaboration \cite{Aktas:2005ty} data, taken at  $W = 82\,\textrm{GeV}$ and $Q^2 = 8\,\textrm{GeV}^2$, with $\delta$ and $n$ respectively fixed to values $0.75$ and $1.54$ \cite{Guzey:2005ec}.
%
%We can determine a formula to find the normalization factors. Indeed, if we introduce the $\varepsilon_{_{Q^2}}$ factor for fixed values of $Q^2$, i.e. for $W$ variable, and the $\varepsilon_{_W}$ factor for fixed values of $W$, i.e. for $Q^2$ variable, by Eq.~(\ref{eq:dipendenza-Wdelta-Q2}) we have:
In this case, the %procedure to rescale the experimental data % basically %consists in normalizing
%is to normalize the experimental data collected by the ZEUS Colla\-boration to those collected by the H1 Collaboration through the adoption of appropriate factors%for normalization
procedure for rescaling depends on the definition of appropriate normalization factors, which here are indicated with $\varepsilon$.
%Using the %trend
%formula given by
%Eq.~(\ref{eq:dipendenza-Wdelta-Q2}), we can determine a %relationship
%formula %in which there are the normalizations for fixed $Q^2$ ($W$ variable), $\varepsilon_{_{Q^2}}$, and for fixed $W$ ($Q^2$ variable), $\varepsilon_{_W}$:
%to find $\varepsilon_{_{Q^2}}$, that represents the normalization factor for fixed values of $Q^2$, i.e. for $W$ variable, and $\varepsilon_{_W}$, that represents the normalization factor for fixed values of $W$, i.e. for $Q^2$ variable:
%
In particular, $\varepsilon_{_{Q^2}}$ represents the normalization factor when we consider the cross section, $\sigma (W)$, as a function of $W$ with fixed $Q^2$; $\varepsilon_{_{W}}$ represents the normalization factor when we consider the cross section, $\sigma (Q^2)$, as a function of $Q^2$ with fixed $W$.
%
%\begin{equation}
%\label{eq:relazione-sezioni-dr-s-r}
%			\sigma_{{r}} (Q^2,W) = \varepsilon_{_{W}} \, \varepsilon_{_{Q^2}} \, \sigma_{{dr}} (Q^2,W) \,,
%\end{equation}
%
%
%\begin{equation}
%\label{eq:relazione-fattori-normalizzazione}
%			\varepsilon_{_{W}} \, \varepsilon_{_{Q^2}} = %\frac{\sigma_{{s}}(Q_{s}^2,W_s)}{\sigma_{{dr}}(Q_{dr}^2,W_{dr})} =
%\frac{(W_s)^{\delta_s}}{(W_{dr})^{\delta_{dr}}} \, \frac{(Q^2_{dr})^{n_{dr}}}{(Q^2_s)^{n_{s}}} \,;%,
%\end{equation}
%
%
By using the Eq.~(\ref{eq:dipendenza-Wdelta-Q2}) we have
\begin{equation}
\label{eq:relazione-sezioni-dr-s-r-Q2-dr}
			\sigma_{{r}} (W) = \frac{\sigma_{{s}} (Q^2_{s},W)}{\sigma_{{dr}} (Q^2_{dr},W)} \,\sigma_{{dr}} (W) = \frac{(Q^2_{dr})^{n_{dr}}}{(Q^2_s)^{n_{s}}} \sigma_{{dr}} (W)\,,
\end{equation}
\begin{equation}
\label{eq:relazione-sezioni-dr-s-r-W-dr}
			\sigma_{{r}} (Q^2) = \frac{\sigma_{{s}} (Q^2,W_s)}{\sigma_{{dr}} (Q^2,W_{dr})} \,\sigma_{{r}} (Q^2) = \frac{(W_s)^{\delta_s}}{(W_{dr})^{\delta_{dr}}} \,\sigma_{{r}} (Q^2)\,,
\end{equation}
%
%and we can determine the formulas to find $\varepsilon_{_{Q^2}}$ and $\varepsilon_{_{W}}$:   allow us to obtain the
where the subscripts ``$dr$", ``$s$" and ``$r$" respectively denote the data to be rescaled, those consi\-dered in %own
scale and %those
the %
rescaled %
data.
The previous relations allow us to obtain the following formulas:
\begin{equation}
\label{eq:definizione_varepsilon_Q2}
			 \varepsilon_{_{Q^2}} = \frac{(Q^2_{dr})^{n_{dr}}}{(Q^2_s)^{n_{s}}} \,,
\end{equation}
\begin{equation}
%\label{eq:definizione_varepsilon_W}
             \varepsilon_{_{W}} = \frac{(W_s)^{\delta_s}}{(W_{dr})^{\delta_{dr}}}\,.
\end{equation}
%
%where the subscripts ``$dr$", ``$s$" and ``$r$" respectively denote the data to be rescaled, those consi\-dered in %own
%scale and %those
%the %
%rescaled %
%data. %So, assuming
In the ``standard'' procedure
the equalities
$\delta_{dr}=\delta_{s}=\delta$ and $n_{dr}=n_s=n$ %.
are %usually
considered valid \cite{Guzey:2005ec}, whereby
the normalization factor $\varepsilon_{_{Q^2}}$, for $\sigma_{dr} (W) \to \sigma_{r} (W)$, is given by the %following
ratio between  $(Q_{dr}^2)^{n}$ and $(Q_{s}^2)^{n}$
%
%\begin{equation}
%\label{eq:definizione_varepsilon_Q2}
%			\varepsilon_{_{Q^2}} = \frac{(Q_{dr}^2)^{n}}{(Q_{s}^2)^{n}} \,,%,
%\end{equation}
%
and the normalization factor $\varepsilon_{_W}$, for $\sigma_{dr} (Q^2) \to \sigma_{r} (Q^2)$, is given by the %following
ratio between $(W_{s})^{\delta}$ and $(W_{dr})^{\delta}$.
%
%\begin{equation}
%\label{eq:definizione_varepsilon_W}
%			\varepsilon_{_W} = \frac{(W_{s})^{\delta}}{(W_{dr})^{\delta}} \,,%,
%\end{equation}
%
%Taking
If we consider $\delta = 0.77$ and $n = 1.54$ \cite{Aktas:2005ty}, the ZEUS cross sections are rescaled to H1 ones through following expressions:
\begin{equation}
\label{eq:varepsilon-Q2}
	\sigma_{r} (W) = \varepsilon_{_{Q^2}} \, \sigma_{dr} (W) \simeq 1.3242 \, \sigma_{dr} (W) \,,
\end{equation}
\begin{equation}
\label{eq:varepsilon-W}
	\sigma_{r} (Q^2) = \varepsilon_{_{W}} \, \sigma_{dr} (Q^2) \simeq 0.9389 \, \sigma_{dr} (Q^2) \,.
\end{equation}
%
%In Fig.~\ref{fig:Ciappetta_DVCS2011-1_fig1}
%we %can appreciate graphically
%graphically illustrate the effect of the procedure for rescaling the data. %where
%The trend due to the series of rescaled ZEUS data is systematically moved over the trend given by the series of H1 data. % collected by H1.%, although it follows.
%
%
As shown in Fig.~\ref{fig:Ciappetta_DVCS2011-1_fig1}, where we illustrate the effect of the procedure for rescaling the cross section as function of $Q^2$, the rescaled ZEUS data are roughly moved over the H1 data.
To highlight this feature,  %
%after choosing an appropriate curve fit model, %
%two independent fits were performed on two data sets, so that each series is associated with a curve interpolating the data points;  %(brown for H1, green for ZEUS);
%we %can, in this way, %consider
%compare, %
%in this way, %
%the cha\-racteristic behavior of two data sets.
%
we fit lines to data so that it catches the general trend of the two data series and we can compare the trends of two data sets.
\section{\label{Procedure}New procedure}
\begin{figure}
    \center
    \includegraphics[width=\columnwidth]{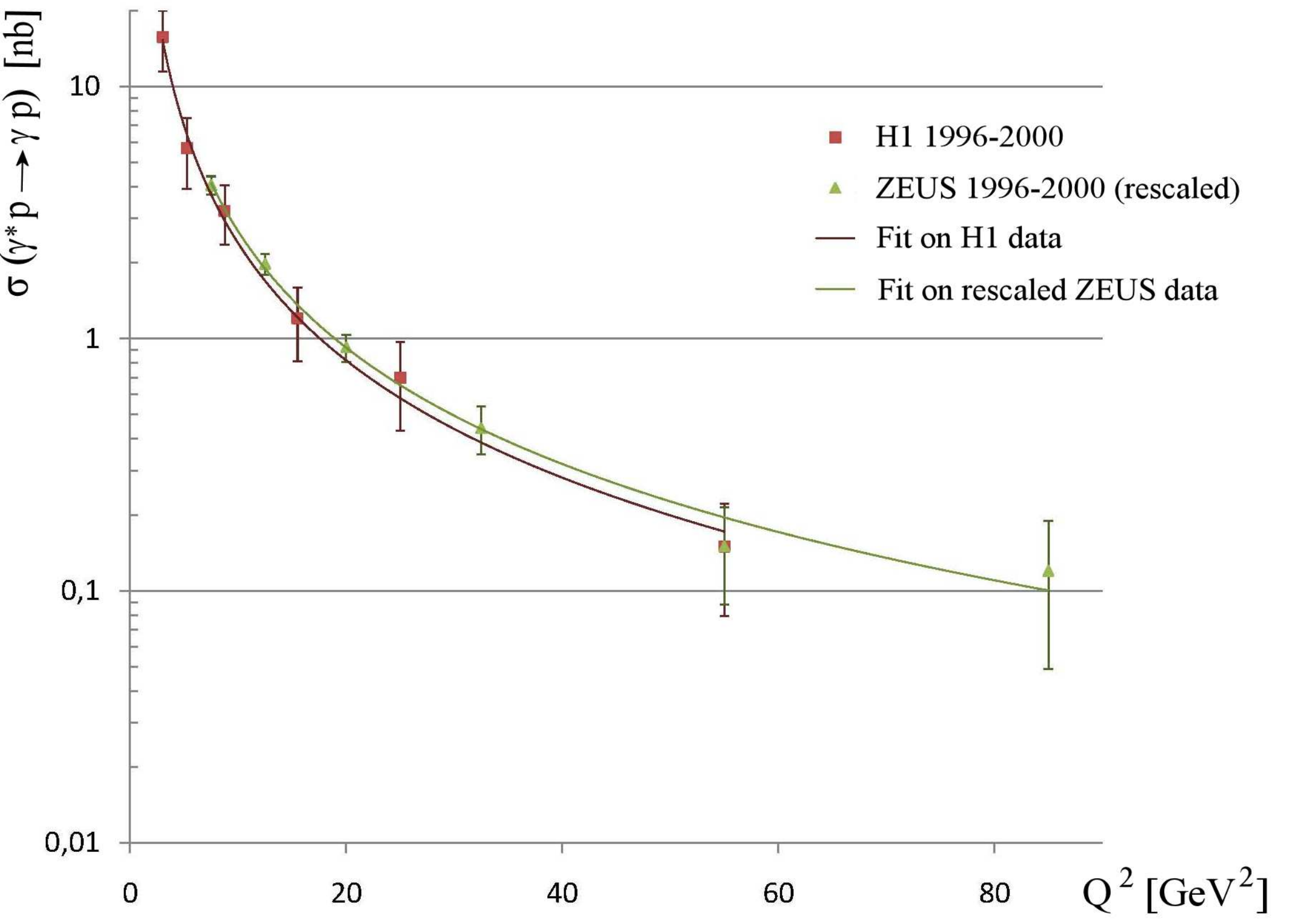}
    \caption{\label{fig:Ciappetta_DVCS2011-1_fig2}
%(Color on the Web only) DVCS cross section $\sigma(\gamma^\ast p \longrightarrow \gamma p$) as function of $Q^2$ for $W=82\,\textrm{GeV}$ ($|t| < 1.0 \,\textrm{GeV}^2$). The error bars re\-present the statistical and systematic uncertainties added in quadrature. The experimental data collected by the ZEUS Collaboration \cite{Chekanov:2003ya} have been rescaled to those collected by the H1 Collaboration \cite{Aktas:2005ty} using the normalization factor $\varsigma_{_W} = 3 / 4 = 0.75$ determined by Eq.~(\ref{eq:costante-normalizzazione-Q2=55GeV}).
}
\end{figure}

\begin{figure}
    \center
    \includegraphics[width=\columnwidth]{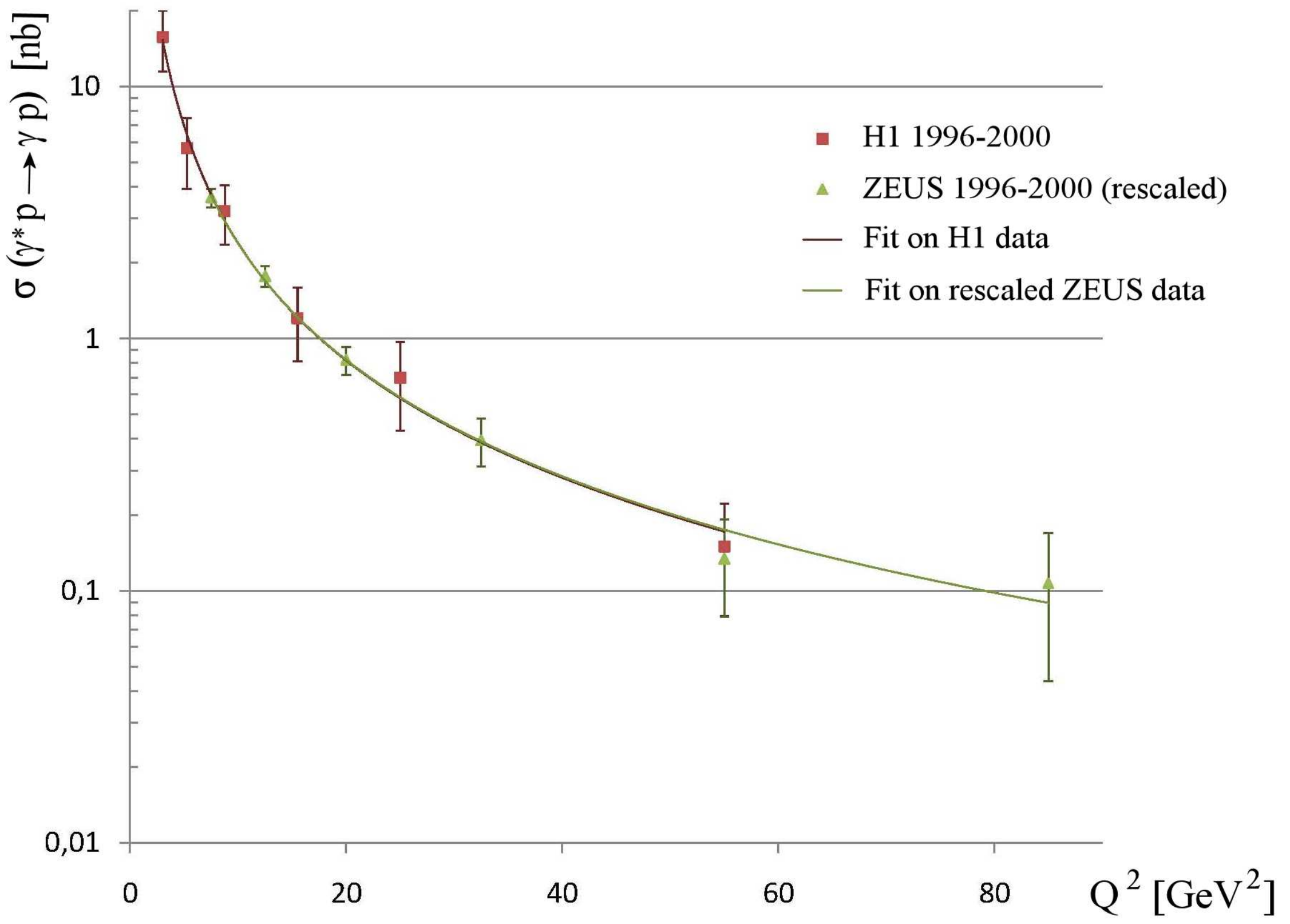}
    \caption{\label{fig:Ciappetta_DVCS2011-1_fig3}
%(Color on the Web only) DVCS cross section $\sigma(\gamma^\ast p \longrightarrow \gamma p$) as function of $Q^2$ for $W=82\,\textrm{GeV}$ ($|t| < 1.0 \,\textrm{GeV}^2$). The error bars re\-present the statistical and systematic uncertainties added in quadrature. The experimental data collected by the ZEUS Collaboration \cite{Chekanov:2003ya} have been rescaled to those collected by the H1 Collaboration \cite{Aktas:2005ty} using the normalization factor $\zeta_{_W} = 0.67$. % determined by Eq.~(\ref{eq:costante-normalizzazione-zeta}).
}
\end{figure}

%
%In
%Fig.~\ref{fig:Ciappetta_DVCS2011-1_fig1} %
%shows %
%the fits performed on the series of data %collected %(
%and
%rescaled %)
%and collected
%from the two Colla\-borations. %show clearly that %, in practice,
%Clearly it seems that %the normalization standard procedure, achieved through Eq.~(\ref{eq:varepsilon-W}),
%
From analysis of the two fits shown in Fig.~\ref{fig:Ciappetta_DVCS2011-1_fig1}, it is found that the rescaled ZEUS data tend to remain higher than those of H1; therefore it seems that
the ``standard'' procedure above described does not rescale
%does not adapt
the %rescaled
ZEUS experimental %points
data to those of H1. % in its scale.
In particular, data %
points %
for $Q^2 = 55\,\textrm{GeV}^2$ are not superimposed,
although they are consistent within the error bars. %Indeed, we would expect that the %overall
%data, %(i.e. the
%rescaled and in scale ones, %)
%are
Indeed, we would expect that, after rescaling, the data will be superimposed when they refer to the same value of $Q^2$. %In this way,
%Therefore, we might thought to improve the rescaling procedure
In this regard, we might consider an alternative rescaling procedure
by normalizing the ZEUS data to those of H1 %,  %with a different method,
and using the following normalization factor:
\begin{equation}
\label{eq:costante-normalizzazione-Q2=55GeV}
			\varsigma_{_W} = \frac{\sigma_{s}(Q^2=55\,\textrm{GeV}^2)}{\sigma_{dr}(Q^2=55\,\textrm{GeV}^2)} = \frac{0.15}{0.20} = \frac{3}{4} \,,
\end{equation}
where $0.15$ and $0.20$ are the cross section values measured by ZEUS and H1 experiments at the same value of $Q^2$.
%
%
%Figure~\ref{fig:Ciappetta_DVCS2011-1_fig2}
%shows the experimental data of H1 and those of ZEUS rescaled according to norma\-lization given by Eq.~(\ref{eq:costante-normalizzazione-Q2=55GeV}).
%
Fig.~\ref{fig:Ciappetta_DVCS2011-1_fig2}
shows the ZEUS data rescaled according to Eq.~(\ref{eq:costante-normalizzazione-Q2=55GeV}).
%
%    \includegraphics[width=\columnwidth]{Ciappetta_DVCS2011-1_fig2-eps-converted-to.pdf}
%
%As it is noted,
As seen in the previous figures, %in the approximation of rescaled ZEUS data trend to that of H1,
the changing of normalization factor %from the value 0.9389 ($= \varepsilon_{_{W}}$) to that of 0.75 ($= \varsigma_{_{W}}$)
has
%produced a good improvement,
given a better approximation of rescaled ZEUS data to those of H1.
However, from Fig.~\ref{fig:Ciappetta_DVCS2011-1_fig2} it %seems
is clear that it is not possible to conduct statistical analysis on overall data.
% but not still functional in order to give possible statistical analysis of overall data.
%%%%%%%%%%
%In the current study we suggest that not the two data sets but the trends, determined by independent fits on H1 and rescaled ZEUS data, must overlap, i.e. the characteristic parameters of trends must have similar values.
In the current study we suggest that the rescaling %correct
procedure %provides
should be based on a a comparison of the trend determined by fit %on
to the rescaled ZEUS data %and
with the trend determined by fit %on
to the H1 data, i.e. the characteristic parameters of both fits must have similar values %and the curves must overlap significantly.
since the fitting curves must be close to each other.
This observation is %right
physically correct %
%if we think that
because the process is the same for both %ZEUS and H1
Collaborations, although the data are collected at different $Q^2$ and $W$ %
values. In effect, if %
the ZEUS and H1 data were taken at the same energies, we would expect similar values %of
for the characteristic parameters of %data
the fits. This consideration is the basis %for
of any rescaling procedure.
%%%%%%%%%%
%Note that,
Also to avoid experimenter's bias, we suggest to consider the trend of the fits to the data rather than data points itself.
%%%%%%%%%%
Therefore it is ne\-cessary to redefine another normalization factor, %which will be determined by varying the value of $\varsigma_{_{W}}$, given in Eq.~(\ref{eq:costante-normalizzazione-Q2=55GeV}), up to a value $\zeta_{_{W}}$ for which the trends of data (i.e. those in their scale and rescaled ones) of the two Collaborations identify the same curve:
which we indicate with $\zeta_{_{W}}$. The latter can be determined %for attempts
by varying %further
the value of $\varsigma_{_{W}}$ until there is good agreement on characteristic parameters of fits, as previously highlighted.
%
%\begin{equation}
%\label{eq:costante-normalizzazione-zeta}
%	\sigma_{r} (Q^2) = \zeta_{_W} \, \sigma_{dr} (Q^2) \,.
%\end{equation}
%
So we find $\zeta_{_W} = 0.67$, value for which the parameters of fits describe the same curve as the Fig.~\ref{fig:Ciappetta_DVCS2011-1_fig3} shows.
%
%Figure~\ref{fig:Ciappetta_DVCS2011-1_fig3}
%shows the two sets of data, H1 and rescaled ZEUS as function of $\zeta_{_W}$ (for $\zeta_{_W} = 0.67$), and the curves determined by the fit on each data set. In this case% it is noted that
%, the two trends %are
%seem almost perfectly adapted to each other.
%
%    \includegraphics[width=\columnwidth]{Ciappetta_DVCS2011-1_fig3-eps-converted-to.pdf}
%
In this case, %combining the two sets of data, %i.e. those of H1 and rescaled ZEUS with normalization factor $\zeta_{_W} = 0.67$, we note that
the fit on overall data gives a $n$-value compatible with that %taken
obtained by the H1 Collaboration, %\cite{Aktas:2005ty}, %
i.e. $n = 1.54 \pm 0.09 \pm 0.04$ \cite{Aktas:2005ty}, where the first error is statistical, the second systematic.
\section{\label{new-normalization-factor-fixed-W}New normalization factor for fixed~$W$}
Adopting the following power-type function
\begin{equation}
\label{eq:sigma-function-Q2-type-power}
	\sigma (Q^2)=a  \times \left[ 1/Q^2 \right]^{n} \,
\end{equation}
and %
by %
performing a fit on H1 data, we have $a_s=83.47 \pm 10.96$ and $n_s = 1.54 \pm 0.06$, with reduced chi-square $\chi^2 / \textrm{d.o.f.} = 0.15$; these parameters are compa\-tible with those %determined
calculated %
by performing a fit on the rescaled ZEUS data using $\zeta_{_W}$ % = 0.67$
factor: $a_{\zeta_{_W}} = 80.99 \pm 8.71$ and $n_{r} = 1.53 \pm 0.04$, with $\chi^2 / \textrm{d.o.f.} = 0.26$. %Moreover, %the multiplicative constant obtained by fit on standard rescaled ZEUS data (i.e. using $\varepsilon_{_W} = 0.9389$ factor) is $a_{\varepsilon_{_W}} = 113.50 \pm 12.21$,
Furthermore, if we use the factor $\varepsilon_{_W}$ for the rescaling procedure, the fit on %standard
rescaled ZEUS data %
gives %
$a_{\varepsilon_{_W}} = 113.50 \pm 12.21$, %
%%%
which is inconsistent with $a_s$. Hence we must introduce the factor $\zeta_{_W}$ and reject the standard procedure. % by which we define $\varepsilon_{_W}$.
%%%
It's possible to move from $\varepsilon_{_W}$ to $\zeta_{_W}$ %factor
by applying the following formula:
\begin{equation}
\label{eq:relazione-zeta-varsigma}
		\zeta_{_W} = \frac{a_{\zeta_{_W}}}{a_{\varepsilon_{_W}}} \, \varepsilon_{_W} \equiv \Xi_{_W} \, \varepsilon_{_W} \,,
\end{equation}
%
%Therefore, if we want to rescale the ZEUS data \cite{Chekanov:2003ya} in a functional way such as we can infer a statistical analysis on them and on H1 ones \cite{Aktas:2005ty}, represented in its scale, it is necessary to multiply the $\varepsilon_{_W}$ normalization factor, given by Eq.~(\ref{eq:varepsilon-W}), for the $\Xi_{_W}$ factor.
where we introduce the factor $\Xi_{_W}$. %
This one may show a %further possible dependence of the normalization factor by $W$ energy and the dependence could not be considered taking the only $\varepsilon_{_W}$.
$W$ dependence which could not be considered taking only the factor $\varepsilon_{_W}$.
The %$\Xi_{_W}$ factor value is found from Eq.~(\ref{eq:relazione-zeta-varsigma}):
value of $\Xi_{_W}$ is found from Eq.~(\ref{eq:relazione-zeta-varsigma}):
\begin{equation}
\label{eq:relazione-zeta-varsigma-Xi}
	\Xi_{_W} = \frac{\zeta_{_W}}{\varepsilon_{_W}} \simeq 0.71 \,.
\end{equation}
%
%In particular,
We might ask if $\Xi_{_W}$
%factor $\Xi_{_W}$
can be determined using the ratio between the $W$ ener\-gies with which the H1 and ZEUS Collaborations have performed their measurements. Actually this event happens when we raise the ratio to the fourth power:
\begin{equation}
\label{eq:relazione-Xi-come-rapporto-energie}
	 \Xi_{W} \simeq \left( \frac{W_{s}}{W_{dr}} \right)^{4} = \left( \frac{82}{89} \right)^{4} = 0.72 \,.
\end{equation}
%
%So, in order to adapt the data of a Collaboration with the other one, it appears that it is necessary to introduce the following normalization factor:
%
Hence, in order to make %efficient
the rescaling procedure
more efficient in statistical terms, it is necessary to replace $\varepsilon_{_W}$ with the following normalization factor:
\begin{equation}
\label{eq:nuova-costante-normalizzazione-dati-epsilon-W}
	\varepsilon_{_W}^{\,'} \simeq \left( \frac{W_{s}}{W_{dr}} \right)^{4+\delta} = \left( \frac{82}{89} \right)^{4 + \delta} = 0.6766 \,,
\end{equation}
where we use $\delta = 0.77$ % \pm 0.23 \pm 0.19$
\cite{Aktas:2005ty}.
Since the %value of %new normalization factor
$\varepsilon_{_W}^{\,'}$ value
is approximately equal to $\zeta_{_W}$, %previously determined and in Fig.~\ref{fig:Ciappetta_DVCS2011-1_fig3}
%the effect of the new normalization for the used experimental data is well reproduced.
%and so
the curve %reproduced
in Fig.~\ref{fig:Ciappetta_DVCS2011-1_fig3} represents approximately the fit
of the power function
to ZEUS data rescaled by the $\varepsilon_{_W}^{\,'}$ factor. % of Eq.~\ref{eq:nuova-costante-normalizzazione-dati-epsilon-W}.
%In this case, if we run a fit on the overall data, i.e. H1 data with ZEUS data rescaled using $\varepsilon_{_W}^{\,'}$ normalization factor,
If we fit the overall data %with
using the function of Eq.~\ref{eq:sigma-function-Q2-type-power}, %
we obtain $a=84.21 \pm 9.06$ and $n = 1.54 \pm 0.04$, with $\chi^2 / \textrm{d.o.f.} = 0.26$; these parameters are clearly compatible with
those
%the parameters $a_s$ and $n_s$
obtained by the fit %performed only on
to the H1 data.
%%%
%Note that
%
%
\section{\label{method-rescaling}Method for rescaling the DVCS data collected at different energies}
Introducing the
following
function
\begin{equation}
\label{eq:dipendenza_P}
	\mathcal{P} (Q, W) = W^{4+\delta} \times \left( \frac{1}{Q^2} \right)^{n} \,,
\end{equation}
Eq.~(\ref{eq:dipendenza-Wdelta-Q2}) can be written as
\begin{equation}
\label{eq:introduzione_P}
	\sigma%_{_{DVCS}}
                         (Q^2,W) \propto \frac{1}{W^4} \, \mathcal{P} (Q, W) \,,
\end{equation}
%
%where
%
%\begin{equation}
%\label{eq:dipendenza_P}
%	\mathcal{P} (Q, W) = W^{4+\delta} \times \left( \frac{1}{Q^2} \right)^{n} \,,
%\end{equation}
%
%thus,
whereby, according %the proposed rescaling,
the rescaling procedure here proposed, we have to carry out the ratio between the quantities $\mathcal{P}_s$ %of data in its scale
and $\mathcal{P}_{dr}$: %, of data to be rescaled:
%
%\begin{equation}
%\label{eq:relazione-P-dr-s}
%			\sigma_{{r}} (Q^2,W) = \frac{\mathcal{P}_{s} (Q_{s},W_{s})}{\mathcal{P}_{dr} (Q_{dr},W_{dr})} \, \sigma_{{dr}} (Q^2,W) \,.
%\end{equation}
%
%
%
%
\begin{equation}
\label{eq:relazione-P-dr-s-sigmaW}
			\sigma_{{r}} (W) = \frac{\mathcal{P}_{s} (Q_{s},W)}{\mathcal{P}_{dr} (Q_{dr},W)} \, \sigma_{{dr}} (W) \,,
\end{equation}
\begin{equation}
\label{eq:relazione-P-dr-s-sigmaQ2}
			\sigma_{{r}} (Q^2) = \frac{\mathcal{P}_{s} (Q,W_{s})}{\mathcal{P}_{dr} (Q,W_{dr})} \, \sigma_{{dr}} (Q^2) \,.
\end{equation}
In general, the differential cross section for the DVCS process, $d\sigma/dt$, can be expressed at high energies \cite{Barone:2002cv} as
\begin{equation}
\label{eq:sezione-urto-differenziale-teorica}
	\frac{d\sigma}{dt} = \frac{1}{16\,\pi \,s^2} \, |\mathcal{M}|^2 \,,
\end{equation}
where the variable $t$ is the square of the four-momentum transferred at the proton vertex, $s$ is the squared centre-of-mass energy of
the incoming system, i.e. $s=W^2$, and $\mathcal{M}$ is the DVCS amplitude. If the $W$ dependence of the integrated cross section $\int\left( d\sigma/dt \right) dt$ is the same, over the relevant $W$ domain, as the $W$ dependence of the differential cross section\footnote{See footnote 14 of Ref. \cite{Adloff:1999kg}.} $d\sigma/dt$ for $t=\langle t \rangle$, then the cross section can be expressed, as indicated in Eq.~\ref{eq:introduzione_P}, in terms of $W^{-4}$. These considerations suggest that the function $\mathcal{P}$ is proportional to the integrated squared modulus of the DVCS amplitude. Therefore, according the Eq.~\ref{eq:relazione-P-dr-s-sigmaW} and Eq.~\ref{eq:relazione-P-dr-s-sigmaQ2}, the ZEUS measurements can be rescaled to the values of the H1 measurements by performing the ratio between the integrated squared modula of scattering amplitudes of the process studied in H1 and ZEUS experiments. Thus, it is interesting to note that the rescaling procedure depends essentially on the scattering amplitudes and that these latter contain all the information about the dynamics of the process.
%
%%%
%
%
\section{\label{parametro-delta}Dependence of $\delta$ parameter on $Q^2$}
\begin{figure}
    \center
    \includegraphics[width=\columnwidth]{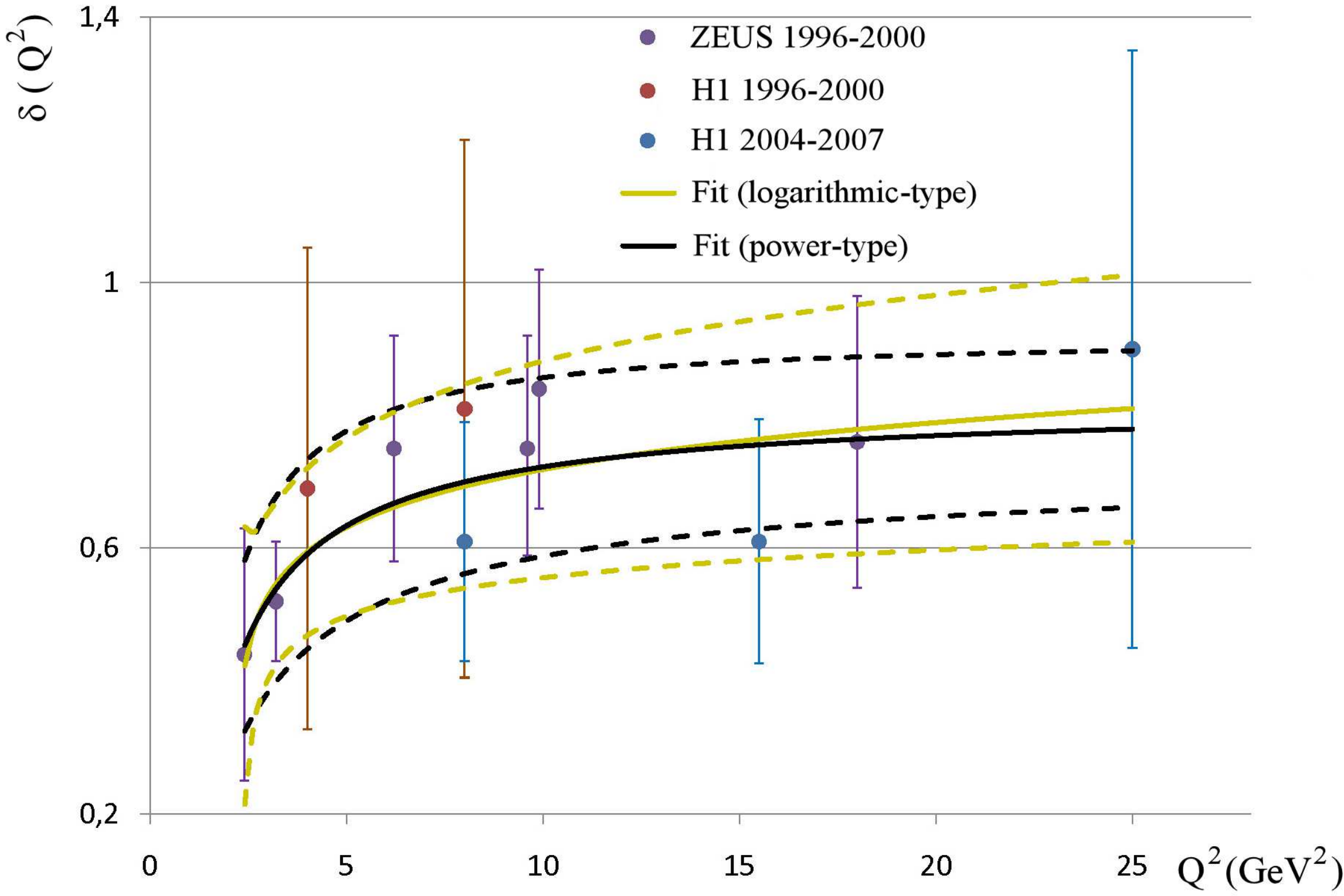}
    \caption{\label{fig:Ciappetta_DVCS2011-1_fig4}
%(Color on the Web only) $\delta$ parameter as a function of $Q^2$. The experimental values are given in Table~\ref{Tab:Ciappetta_DVCS2011-1_Table}. Performed fits show probable trends. The dotted lines indicate the errors associated with each curve. % (red for logarithmic-type, black for power-type fit).
}
\end{figure}

%
%In the reference \cite{Aktas:2005ty} %there are
%it is reported a list of %
%the
%$\delta$-values calculated for various $Q^2$. These values are different, although each one occurs in the error bars of the other. In light of this eventuality, we consider the parameter value supplied in some references \cite{Chekanov:2003ya,Chekanov:2008vy,:2009vda} in order to verify a possible behavior of $\delta$ as function of $Q^2$. % (see Table~\ref{Tab:Ciappetta_DVCS2011-1_Table}).
%
In Table~\ref{Tab:Ciappetta_DVCS2011-1_Table} we have collected %some
the $\delta$-values calculated by ZEUS and H1 experiments \cite{Aktas:2005ty,Chekanov:2003ya,Chekanov:2008vy,:2009vda}.
%
% {Tab:Ciappetta_DVCS2011-1_Table}
%
Taking account that %some
several values ​​do not lie within the error bars of other values, we consider the possibility of treating the $\delta$-parameter as function of $Q^2$, contrary to what is given in literature which states that $\delta$ is independent of $Q^2$ within the errors \cite{:2009vda}.
%
%Using several different functions
All the functions used to fit data of Table~\ref{Tab:Ciappetta_DVCS2011-1_Table} exhibit a similar trend\footnote{\label{OriginPro8}Through \textit{OriginPro 8} we identified %as many as
23 %different
functions able to fit the experimental %points
data of Table~\ref{Tab:Ciappetta_DVCS2011-1_Table} %
with a %low value of the
reduced chi-square value ranging from %a mi\-nimum of
$\sim 0.2$ to %a maximum of
$\sim 0.6$. Here we want to show that $\delta$ is dependent by $Q^2$; nevertheless, we are not so interested to %statistical analysis of parameters characte\-rizing the different functions obtained by the fits,
statistically analyze the data, but our interest is to check, also by `eye', %if
whether %this
the dependence is actual.}
%to fit the %experimental points
%data of Table~\ref{Tab:Ciappetta_DVCS2011-1_Table}, we note that all these functions exhibit a similar trend,
especially for low values of $Q^2$.
In Fig.~\ref{fig:Ciappetta_DVCS2011-1_fig4} %
two %curves
fits %
are shown:
one %
is %
logarithmic-type, %(red curve)
%and the other
another one is %
power-type. %(black curve).
The logarithmic-type curve is given by the following equation:
\begin{equation}
\label{eq:andamento-delta-da-fit}
	\delta(Q^2)=\delta_0 - \delta_1 \ln (Q^2 + \delta_2) \,,
\end{equation}
where $\delta_0 = 0.5421 \pm 0.0768$, $\delta_1 = -0.0857 \pm 0.0389$ and $\delta_2=-2.1511 \pm 0.5414$, %
with %(
$\chi^2/\textrm{d.}\textrm{o.}\textrm{f.} = 0.2497$; the power-type curve is given by the following equation:
\begin{equation}
\label{eq:andamento-delta-da-fit-tipopotenza}
	\delta(Q^2)=\delta' \, \left[1 - (Q^2)^{-\delta_p} \right] \,,
\end{equation}
where $\delta' = 0.8232 \pm 0.0887$ and $\delta_p = 0.9137 \pm 0.2455$, %(
with $\chi^2/\textrm{d.}\textrm{o.}\textrm{f.} = 0.2051$. %).
%Because
Since $\delta$ is treated as a function of $Q^2$, we have that %even
the %normalization
factor $\varepsilon_{_W}^{\,'}$ %, given by Eq.~(\ref{eq:nuova-costante-normalizzazione-dati-epsilon-W}),
depends on $Q^2$:
\begin{equation}
\label{eq:costante-normalizzazione-dati-epsilon-W-definitiva}
	\varepsilon_{_W}^{\,'} = \left( \frac{W_{s}}{W_{dr}} \right)^{4+\delta(Q^2)}  \,.
\end{equation}
Figure~\ref{fig:Ciappetta_DVCS2011-1_fig5}
shows the %trends determined by independent fits on H1 and rescaled ZEUS data %, where factor, given
%obtained by Eq.~(\ref{eq:costante-normalizzazione-dati-epsilon-W-definitiva}) %,
%and Eq.~(\ref{eq:andamento-delta-da-fit}). % are used.
%
trend determined by fit to the ZEUS data,
which are rescaled by using Eq.~(\ref{eq:costante-normalizzazione-dati-epsilon-W-definitiva}) and Eq.~(\ref{eq:andamento-delta-da-fit}).
%
%As we can see, the trends overlap% perfectly.
%
This trend is superimposed to that determined by fit to the H1 data. %
In effect, %
%
%$a_s$ and $n_s$ are compatible with
%
%the parameters obtained by fit on rescaled ZEUS data %
%that are $a_{r}=83.74 \pm 8.99$ and $n_r = 1.54 \pm 0.04$ ($\chi^2 / \textrm{d.o.f.} = 0.26$). %, which are compatible with $a_s$ and $n_s$.
by performing a fit to the rescaled ZEUS data, we have $a_{r}=83.74 \pm 8.99$ and $n_r = 1.54 \pm 0.04$, with $\chi^2 / \textrm{d.o.f.} = 0.26$; these parameters are compatible with $a_s$ and $n_s$ obtained by performing a fit to the H1 data.
%It is interesting to note that the trends %continue to
%overlap in spite of an additional dependence by $Q^2$, introduced by the $\delta$ parameter in an independent manner from rescaling analysis conducted in Sec.~\ref{new-normalization-factor-fixed-W}.
It is interesting to note that the trends overlap although %the parameter $\delta$
the visible dependence of $\delta$ on $Q^2$
has introduced a %further
dependence of $\varepsilon_{_W}^{\,'}$ on $Q^2$  in an independent manner with respect to the rescaling analysis conducted in Sec.~\ref{new-normalization-factor-fixed-W}.
Clearly, the growth, %of $\delta$
at low $Q^2$, and the flattening, at high $Q^2$, %does
of $\delta$
do not fundamentally modify the rescaling procedure %here
proposed in this paper.
\section{Conclusion}
%
%In the light of
Through the analysis conducted here,
%it appears that the best way to rescale ZEUS data to H1 ones is to adopt, as rescaling factor for fixed $W$, the quantity which is defined with $\varepsilon_{_W}^{\,'}$, given by Eq.~(\ref{eq:costante-normalizzazione-dati-epsilon-W-definitiva}) where the $\delta$ parameter is dependent by $Q^2$.
%
it appears that at fixed $W$ the ZEUS data \cite{Chekanov:2003ya} may be rescaled to H1 ones \cite{Aktas:2005ty} by using $\varepsilon_{_W}^{\,'}$, i.e. by %
applying Eq.~(\ref{eq:costante-normalizzazione-dati-epsilon-W-definitiva}) where the $\delta$-parameter is dependent on $Q^2$.
%
%%%%%%%%%%
%The proposed rescaling gives new results that we will use in forthcoming works, where we will introduce a phenomenological picture for the DVCS process. In effect,
%
%The new procedure gives results that we will use in future works, where we will present a phenomenological model of the DVCS process. In effect,
%
%%%%%%%%%%
%the procedure %used to find a relationship able to rescale the experimental data collected at diffe\-rent $W$ energies
%can be implemented to find %also
%
The Eq.~(\ref{eq:introduzione_P}) shows that the new rescaling procedure is consistent with the theoretical framework, as mentioned in Sec.~\ref{method-rescaling}. Therefore, the procedure may be implemented to find
a suitable formula to rescale the data taken at different $Q^2$. Then it will be useful to consider $\delta$ as a %$Q^2$
function %
of $Q^2$, %as so as to better define its behavior.
in order to better define its behavior.
The results obtained by the adoption of the new rescaling procedures will be collected and used in future works, where we will present a phenomenological model of the DVCS process.
\section*{Acknowledgements}
The results here presented were discussed at the earlier stage of this contribution with R. Fiore who is gratefully acknowledged. Thanks are due to G. Nicastro and V. Pingitore for the help in preparation of the manuscript. %
Special thanks to A. Papa for his invaluable advice and encouragement.
The autor is grateful to his family for moral and financial support during this work.
%



%% References

%
\section*{References}

\bibliographystyle{elsarticle-num}
\bibliography{Ciappetta_DVCS2011-1}

\begin{thebibliography}{1}
\expandafter\ifx\csname url\endcsname\relax
  \def\url#1{\texttt{#1}}\fi
\expandafter\ifx\csname urlprefix\endcsname\relax\def\urlprefix{URL }\fi
\expandafter\ifx\csname href\endcsname\relax
  \def\href#1#2{#2} \def\path#1{#1}\fi

\bibitem{Aktas:2005ty}
A.~Aktas, et~al., {Measurement of deeply virtual Compton scattering at HERA},
  Eur. Phys. J. C 44 (2005) 1--11.
\newblock \href {http://arxiv.org/abs/hep-ex/0505061}
  {\path{arXiv:hep-ex/0505061}}, \href
  {http://dx.doi.org/10.1140/epjc/s2005-02345-3}
  {\path{doi:10.1140/epjc/s2005-02345-3}}.

\bibitem{Chekanov:2003ya}
S.~Chekanov, et~al., {Measurement of deeply virtual Compton scattering at
  HERA}, Phys. Lett. B 573 (2003) 46--62.
\newblock \href {http://arxiv.org/abs/hep-ex/0305028}
  {\path{arXiv:hep-ex/0305028}}, \href
  {http://dx.doi.org/10.1016/j.physletb.2003.08.048}
  {\path{doi:10.1016/j.physletb.2003.08.048}}.

\bibitem{Guzey:2005ec}
V.~Guzey, M.~V. Polyakov, {Dual parameterization of generalized parton
  distributions and description of DVCS data}, Eur. Phys. J. C 46 (2006)
  151--156.
\newblock \href {http://arxiv.org/abs/hep-ph/0507183}
  {\path{arXiv:hep-ph/0507183}}, \href
  {http://dx.doi.org/10.1140/epjc/s2006-02491-0}
  {\path{doi:10.1140/epjc/s2006-02491-0}}.

\bibitem{Barone:2002cv}
V.~Barone, E.~Predazzi, {High-energy particle diffraction}, Berlin, Germany:
  Springer (2002) 407 p.

\bibitem{Adloff:1999kg}
C.~Adloff, et~al., {Elastic electroproduction of rho mesons at HERA}, Eur.
  Phys. J. C 13 (2000) 371--396.
\newblock \href {http://arxiv.org/abs/hep-ex/9902019}
  {\path{arXiv:hep-ex/9902019}}, \href
  {http://dx.doi.org/10.1007/s100520050703} {\path{doi:10.1007/s100520050703}}.

\bibitem{Chekanov:2008vy}
S.~Chekanov, et~al., {A measurement of the $Q^2$, $W$ and $t$ dependences of
  deeply virtual Compton scattering at HERA}, JHEP 05 (2009) 108.
\newblock \href {http://arxiv.org/abs/hep-ex/0812.2517}
  {\path{arXiv:hep-ex/0812.2517}}, \href
  {http://dx.doi.org/10.1088/1126-6708/2009/05/108}
  {\path{doi:10.1088/1126-6708/2009/05/108}}.

\bibitem{:2009vda}
F.~D. Aaron, et~al., {Deeply Virtual Compton Scattering and its Beam Charge
  Asymmetry in $e^\pm p$ Collisions at HERA}, Phys. Lett. B 681 (2009)
  391--399.
\newblock \href {http://arxiv.org/abs/hep-ex/0907.5289}
  {\path{arXiv:hep-ex/0907.5289}}, \href
  {http://dx.doi.org/10.1016/j.physletb.2009.10.035}
  {\path{doi:10.1016/j.physletb.2009.10.035}}.

\end{thebibliography}






% \end{thebibliography}

%
\newpage
\begin{table}[htbp]
\caption{\label{Tab:Ciappetta_DVCS2011-1_Table}{$\delta$-values %supplied
collected by various Collaborations. For each measurement, the first are statistical errors and the second ones are sy\-stematic.\\}}
\centering
\begin{tabular}{ccc}
%\hline\noalign{\smallskip}
\hline
$Q^2 \,\,[\textrm{GeV}^2]$  & $\delta$ & Reference \\
\hline
$2.4$ & $0.44 \pm 0.19$ & ZEUS 1999-2000 \cite{Chekanov:2008vy}  \\
%\hline
$3.2$ & $0.52 \pm 0.09$ & ZEUS 1999-2000 \cite{Chekanov:2008vy}  \\
%\hline
$4$ & $0.69 \pm 0.32 \pm 0.17$ & H1 1996-1997 \cite{Aktas:2005ty} \\
%\hline
$6.2$ & $0.75 \pm 0.17$ & ZEUS 1996-2000 \cite{Chekanov:2008vy}  \\
%\hline
$8$ & $0.81 \pm 0.34 \pm 0.22$ & H1 1999-2000 \cite{Aktas:2005ty} \\
%\hline
$8$ & $0.61 \pm 0.10 \pm 0.15$ & H1 2004-2007 \cite{:2009vda} \\
%\hline
$9.6$ & $0.75 \pm 0.15^{+0.08}_{-0.06}$ & ZEUS 1996-2000 \cite{Chekanov:2003ya}  \\
%\hline
$9.9$ & $0.84 \pm 0.18$ & ZEUS 1996-2000 \cite{Chekanov:2008vy}  \\
%\hline
$15.5$ & $0.61 \pm 0.13 \pm 0.13$ & H1 2004-2007 \cite{:2009vda} \\
%\hline
$18$ & $0.76 \pm 0.22$ & ZEUS 1996-2000 \cite{Chekanov:2008vy}  \\
%\hline
$25$ & $0.90 \pm 0.36 \pm 0.27$ & H1 2004-2007 \cite{:2009vda} \\
\hline
\end{tabular}
\end{table}
\newpage
\section*{Figure Captions}
\subsection*{Figure \ref{fig:Ciappetta_DVCS2011-1_fig1} }%(Color on the Web only)}
DVCS cross section $\sigma(\gamma^\ast p \longrightarrow \gamma p$) as function of $Q^2$ for $W=82\,\textrm{GeV}$ ($|t| < 1.0 \,\textrm{GeV}^2$, where $t$ is the four momentum transfer squared at the proton vertex). The error bars re\-present the statistical and systematic uncertainties added in quadrature. The experimental data collected by the ZEUS Collaboration \cite{Chekanov:2003ya} have been rescaled to those collected by the H1 Collaboration \cite{Aktas:2005ty} %applying
 using Eq.~(\ref{eq:varepsilon-W}), where $\varepsilon_{_{W}} \simeq 0.9389$.
\subsection*{Figure \ref{fig:Ciappetta_DVCS2011-1_fig2} }%(Color on the Web only)}
DVCS cross section $\sigma(\gamma^\ast p \longrightarrow \gamma p$) as function of $Q^2$ for $W=82\,\textrm{GeV}$ ($|t| < 1.0 \,\textrm{GeV}^2$). The error bars re\-present the statistical and systematic uncertainties added in quadrature. The experimental data collected by the ZEUS Collaboration \cite{Chekanov:2003ya} have been rescaled to those collected by the H1 Collaboration \cite{Aktas:2005ty}
using the normalization factor $\varsigma_{_W} = 3 / 4 = 0.75$ determined by Eq.~(\ref{eq:costante-normalizzazione-Q2=55GeV}).
\subsection*{Figure \ref{fig:Ciappetta_DVCS2011-1_fig3} }%(Color on the Web only)}
DVCS cross section $\sigma(\gamma^\ast p \longrightarrow \gamma p$) as function of $Q^2$ for $W=82\,\textrm{GeV}$ ($|t| < 1.0 \,\textrm{GeV}^2$). The error bars re\-present the statistical and systematic uncertainties added in quadrature. The experimental data collected by the ZEUS Collaboration \cite{Chekanov:2003ya} have been rescaled to those collected by the H1 Collaboration \cite{Aktas:2005ty} using the normalization factor $\zeta_{_W} = 0.67$ determined %by Eq.~(\ref{eq:costante-normalizzazione-zeta})
 in Sec.~\ref{Procedure}.
\subsection*{Figure \ref{fig:Ciappetta_DVCS2011-1_fig4} }%(Color on the Web only)}
\begin{figure}
    \center
    \includegraphics[width=\columnwidth]{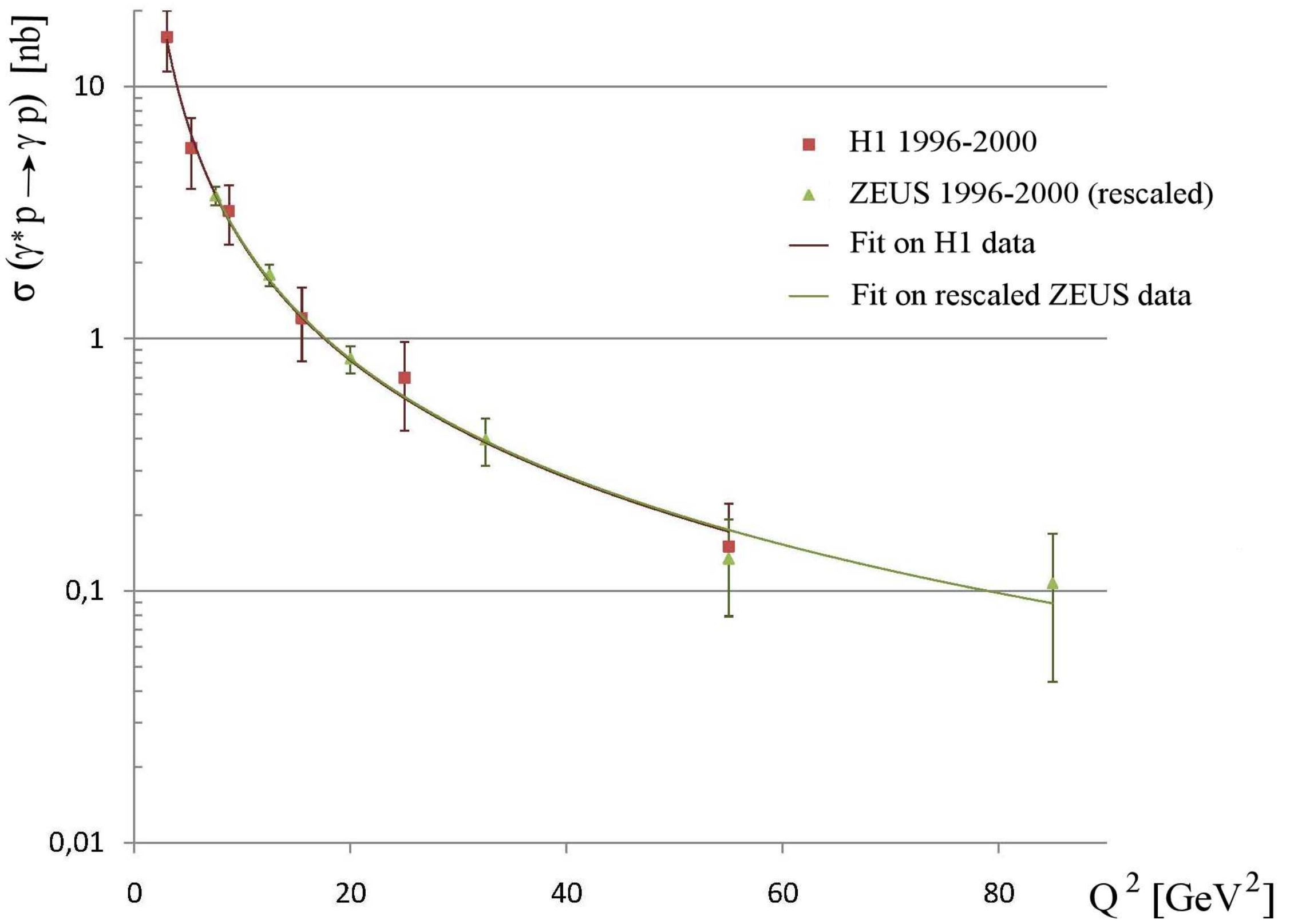}
    \caption{\label{fig:Ciappetta_DVCS2011-1_fig5}
%(Color on the Web only) DVCS cross section $\sigma(\gamma^\ast p \longrightarrow \gamma p$) as function of $Q^2$ for $W=82\,\textrm{GeV}$ ($|t| < 1.0 \,\textrm{GeV}^2$). The error bars re\-present the statistical and systematic uncertainties added in quadrature. The experimental data collected by the ZEUS Collaboration \cite{Chekanov:2003ya} have been rescaled to those collected by the H1 Collaboration \cite{Aktas:2005ty} using Equations (\ref{eq:costante-normalizzazione-dati-epsilon-W-definitiva}) and (\ref{eq:andamento-delta-da-fit}).
}
\end{figure}

$\delta$ parameter as a function of $Q^2$. The experimental values are given in Table~\ref{Tab:Ciappetta_DVCS2011-1_Table}. %Performed fits show probable trends.
Two fits are shown. %
The dotted lines indicate the error bands. % associated with each curve. % (red for logarithmic-type, black for power-type fit).
\subsection*{Figure \ref{fig:Ciappetta_DVCS2011-1_fig5} }%(Color on the Web only)}
DVCS cross section $\sigma(\gamma^\ast p \longrightarrow \gamma p$) as function of $Q^2$ for $W=82\,\textrm{GeV}$ ($|t| < 1.0 \,\textrm{GeV}^2$). The error bars re\-present the statistical and systematic uncertainties added in quadrature. The experimental data collected by the ZEUS Collaboration \cite{Chekanov:2003ya} have been rescaled to those collected by the H1 Collaboration \cite{Aktas:2005ty} using Equations (\ref{eq:costante-normalizzazione-dati-epsilon-W-definitiva}) and (\ref{eq:andamento-delta-da-fit}).%
\newpage

\end{document}